 \documentclass[preprint2]{emulateapj}

\newcommand{\mdot}{M$_{\odot}$~}

\newcommand{\kmsec}{km~s$^{-1}$~}
\newcommand{\htwo}{H$_2$}
\shorttitle{Jet induced star formation at $z>3$}
\shortauthors{Klamer et al.}

\begin{document}

\title{Molecular gas at high redshift: jet-induced star formation?}


\author{I. J. Klamer$^{1,2}$}\email{klamer@physics.usyd.edu.au}
\author{R. D. Ekers$^{2}$}\email{rekers@csiro.au}
\author{E. M. Sadler$^{1}$}\email{ems@physics.usyd.edu.au}
\author{R. W. Hunstead$^{1}$}\email{rwh@physics.usyd.edu.au}
\affil{1. School of Physics A28 University of Sydney, NSW 2006 Australia}
\affil{2. CSIRO Australia Telescope National Facility, PO Box 76 Epping NSW 1710 Australia}


\begin{abstract}
We present an alternative interpretation of the observations of BR~1202$-$0725 at $z=4.695$ and show that its properties are consistent with a relativistic jet, issuing from the quasar core, propagating into the inter-galactic medium and triggering star formation along its path. Prompted by this finding, we reviewed all the $z>3$ objects detected in molecular line emission and found that the distribution of gas and dust in these sources is often spatially or kinematically offset from the host galaxy, and preferentially aligned along the radio axis. These observations suggest to us a scenario in which CO emission observed in high redshift galaxies is located where it forms: along the sites of star formation triggered initially by relativistic jets.
\end{abstract}




\section{Introduction}
\label{intro}
 Ever since the first discoveries that high redshift quasars can be bright sub-millimetre (submm) sources \citep{mcm94} which shine because of dust emission reprocessed from starlight, a large observational effort has been undertaken to find corresponding reservoirs of gas which provide both the fuel and the site for this star formation to occur \citep{omo96a,gui97,gui99}. The gas reservoirs are assumed to be made up almost entirely of molecular hydrogen (\htwo) with trace amounts of heavier molecules like Carbon Monoxide ($^{12}\rm C^{16}\rm O$; hereafter CO). But when and how did these molecules form? At high redshift it remains unclear how to collapse a first generation of metal-free stars and enrich the pristine primordial gas with metals and, eventually, molecules rapidly enough to approach solar metallicity in quasars only 850 million years after the Big Bang \citep{bro04}. \newline
At $z=2$, the co-moving space density of radio galaxies with powerful jets was $1000$ times higher than at the present epoch \citep{wil01}. This hints toward the importance radio jets play in the early universe. Between $2<z<3$ there is a dramatic change in the optical morphology\footnote{It is based on this observation that we chose $z>3$ CO emitters as our redshift cutoff} of the hosts of powerful radio galaxies, from massive ellipticals with old stellar populations to irregular and clumpy distributions aligned along the radio axes \citep{mcc87,cha87,wvb98}. This alignment effect further indicates some connection between the formation of the host galaxy and the collimated jets from the AGN. The mere existence of powerful radio sources residing in such primeval galaxies hints toward a paradigm in which supermassive black holes are in place before their host fully forms. Although the current paradigm for galaxy formation is one in which black holes grow co-evally with host galaxies in hierarchical formation scenarios \citep{whi78,kau93}, each of the above considerations could be consistent within a framework where jet-triggered star formation plays a key role in the formation of some of the first stars and galaxies. In this Letter, we show that the $z=4.695$ quasar BR~1202$-$0725 shows evidence for large scale jet-induced star formation; despite no direct detection of its radio jet. We then review all $z>3$ molecular gas observations to date and show that often the gas and dust is aligned with the radio emission; in the case of bona fide radio galaxies this alignment is along the axes of radio jets. These results have led us to propose a scenario in which CO emission observed at high redshift is located where it formed: along the sites of star formation triggered initially by relativistic jets. We assume a flat $\Lambda$-dominated cosmology with $H_0=71$~\kmsec Mpc$^{-1}$ and $\Omega_{\Lambda}=0.73$. 
\section{BR 1202$-$0725}
\begin{figure}
\includegraphics[width=4cm]{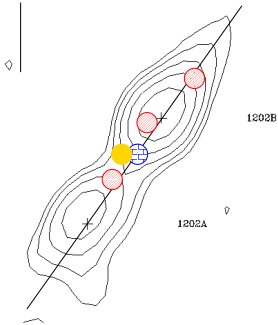}
\caption{\small 1.4~GHz radio continuum contours of BR~1202$-$0725 with pluses marking the CO(2-1) positions (C02). Contour levels are a geometric progression of $\sqrt{2}$ starting at $\pm42.4~\mu$Jy~beam$^{-1}$. Striped, bricked and filled circles depict K-band, I-band and Lyman-$\alpha$ features respectively. North is up and East is to the left. A bar in the top left corner shows the 2$\arcsec$ (13.2 kpc) scale.}
\label{image:cartoon}
\end{figure}

\begin{figure}
\includegraphics[width=7cm]{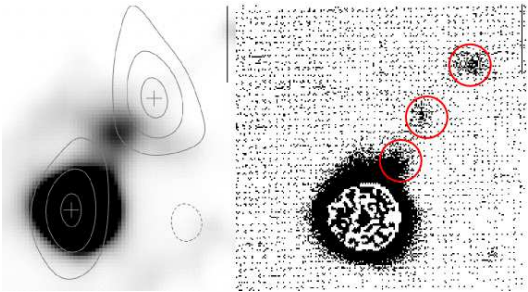}
\caption{\small Multi-wavelength observations of the quasar BR~1202$-$0725. Left: Lyman-$\alpha$ greyscale with CO(2-1) contours overlaid (C02). Right: NICMOS3 K-band emission from H96. The open circles correspond to the K-band features indicated in Figure~\ref{image:cartoon}. North is up and East is to the left. A bar in the top right corner of each panel show the 2$\arcsec$ (13.2 kpc) scale.} 
\label{image:1202}
\end{figure}

At $z=4.695$, BR~1202$-$0725 is an optically bright ($M_B=-28.5$) quasar with broad emission lines \citep{sto96} and the first high redshift quasar confirmed to host massive reservoirs of molecular gas and dust \citep{omo96a,oht96}. Its 1.4 GHz luminosity ($\sim10^{25}$~W~Hz$^{-1}$; \citealp{car02b}; hereafter C02) is consistent with either massive star formation ($\sim2800$~\mdot year$^{-1}$; Equation 21 of \citealp{con92}) or a Fanaroff-Riley type~I radio galaxy \citep{fan74}. Luminous ($\sim10^{45}$~ergs~s$^{-1}$) X-ray emission from the quasar core provides clear evidence that BR~1202$-$0725 plays host to a powerful AGN (I.~J.~Klamer, D.~M.~Alexander \& C.~Vignali, in preparation). This is consistent with \citet{yun00} who measure a radio spectral index\footnote{The spectral index, $\alpha$ (where $S_{\nu}\propto\nu^{\alpha}$) is measured from the integrated emission from BR~1202$-$0725 and its nearby companion between 1.4 and 4.9 GHz. Future high resolution observations should be able to determine $\alpha$ separately for the two components.} of $-0.43\pm0.14$, which is flatter than the narrow range expected for pure star formation: $\langle{\alpha}\rangle=-0.75$, $\sigma_{\alpha}=0.10$ \citep{con83} but consistent with a flat spectrum radio loud AGN. High resolution VLA observations, shown in Figure~\ref{image:cartoon}, show two radio sources separated by about 24 kpc along a position angle (PA) of $-$35$\pm1.0^{\circ}$. The northern source, almost twice as luminous as the quasar at 1.4 GHz, coincides with a dusty, star forming companion (C02) which is undetected at optical wavelengths. If the radio emission from the companion is dominated by star formation, the corresponding star formation rate is a titanic $\sim4400$~\mdot year$^{-1}$; more than three times the rate implied from the submm emission (Equation~26 of \citealt{con92}). However, we find evidence to suggest that there is a relativistic jet emanating from BR~1202$-$0725 along a PA of about $-$35$^{\circ}$. If correct, a portion of the radio emission could be due to synchrotron radiation from the relativistic electrons in the jet and the companion itself the result of jet-induced star formation; hereafter we refer to the quasar as 1202A and the star forming companion as 1202B. This scenario would be somewhat analogous to the 850$\mu$m SCUBA source LE~850.7 associated with jet-induced star formation from the radio lobe of R072 \citep{ivi02}.\newline
Evidence for star formation occurring along a path from 1202A to 1202B has been found in published optical and infra-red observations; these observations are depicted in Figure~\ref{image:cartoon}. Lyman$-\alpha$ emission (left panel; Figure~\ref{image:1202}) from 1202A is extended along a PA of $-$28$\pm5^{\circ}$ with a velocity gradient shifting redward with increasing distance away from the quasar (\citealp{fon98,hu96}; hereafter H96). HST I-band observations (H96) reveal a faint, extended continuum source located about half way between 1202A and 1202B along a position angle (PA) of $-$35$\pm3^{\circ}$ and $0\farcs6$ (4 kpc) from a Lyman$-\alpha$ emitter at $z=4.702$. NICMOS3 K-band observations (right panel; Figure~\ref{image:1202}) of the same field (H96) shows that the quasar is clearly extended along a similar North-West direction. Two more K-band features are also detected, one located just North of 1202B along a PA of $-$36$\pm3^{\circ}$, the other East of 1202B along a PA of $-$30$\pm3^{\circ}$. The presence of such features, without corresponding I-band counterparts, is plausibly due to [O$\textsc{ii}$]$\lambda$3727 line emission, a clear star formation indicator, which falls in the K-band window at this redshift. In addition, the CO line emission is extended and the velocity profiles have structure. VLA imaging (C02) with $2.6''\times1.8''$ resolution reveals unresolved CO(2-1) emission at both 1202A and 1202B, whereas higher resolution observations ($0.3''\times0.21''$) only detect 1202A confidently, suggesting 1202B has been resolved. The CO~line profile at 1202A is both stronger and narrower than at 1202B. Lastly, X-ray emission coincident with 1202B has recently been detected (I.~J.~Klamer, D.~M.~Alexander \& C.~Vignali, in preparation) and is consistent with inverse Compton (IC) radiation against the cosmic microwave background (CMB).\newline Although existing radio observations of 1202A do not have the sensitivity to detect a jet, a scenario in which an inferred jet emanating from 1202A is triggering star formation along its path toward 1202B, is a natural and consistent interpretation of all existing observations. The broad line profile of CO at 1202B would then reflect a bulk, turbulent velocity field associated with star formation from multiple cloud collapse and the gradient in the Lyman-$\alpha$ velocity would be consistent with ionisation along the axis of the jet.
\vspace{-3mm}\section{High redshift CO observations}
\begin{figure}
\includegraphics[width=7cm]{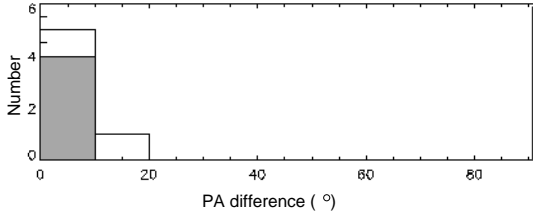}
\caption{\small Stacked histogram of relative differences between the radio-CO (shaded) and radio-dust (unshaded) PAs for the six sources in Table~1 with sufficient PA information. BR~1202$-$0725 and 4C~60.07 have both radio-CO and radio-dust PA information and are counted once in the shaded region. In the absence of an alignment effect, the relative PAs will be randomly distributed into nine bins over $0^{\circ}<$PA$<90^{\circ}$ and the probability that five of the six objects fall in the $0^{\circ}-10^{\circ}$ bin is $1.5\times10^{-5}$.}
\label{image:histo}
\end{figure}

After finding evidence for large scale jet-induced star formation in BR~1202$-$0725, we collected the available data on all  $z>3$ CO emitters to investigate whether this model might be applicable elsewhere. Table~1 presents a brief summary of each source. The optical redshift is listed in Column~2 followed by the CO redshift in Column~3. Columns~4-7 summarise the radio, CO and submm~ information available for each source and the references used in constructing this table are listed in Column~8. Superscripts in Column~1 denote the source classification. Nine out of the twelve known $z>3$ CO emitters show spatial or kinematic structure. Some characteristics are:\newline
\textbf{$\bullet$} The CO and/or dust are spatially extended and aligned with the radio axis (e.g. B3~J2330+3927, 4C~60.07, 6C~1909+722).\newline
\textbf{$\bullet$} The CO profile shows two kinematic and/or spatially distinct components and the host galaxy coincides with one (e.g. BRI~1335$-$0417, BR~1202$-$0725, 4C~60.07).\newline
\textbf{$\bullet$} The CO velocity is offset with respect to the optically derived velocity (e.g. TN~J0121+1320, BR1~0952$-$0115, B3~J2330+3927, 4C~60.07, BR1~1335$-$0417).\newline
\textbf{$\bullet$} The profile of any CO line detected at the position of a host galaxy is narrow compared with much broader CO emission offset from the host (e.g. BR~1202$-$0725, 4C~60.07).\newline
Apart from the bona fide radio galaxies, there is still no direct evidence for jets in the other CO emitters. However, each of these properties is consistent within a framework in which the molecular gas and dust are associated directly with star formation along the axis of an inferred relativistic jet. In this scenario, one would expect that the CO profile from emission outside the host galaxy will be broad reflecting the bulk gaseous velocity field from multiple cloud collapse, that the CO and Lyman-$\alpha$ will be tracing very different physical regions and so the derived redshifts for each will be different, and that the CO and dust will be extended and aligned with observable synchrotron radio emission. If jet-cloud interactions are responsible for CO production at high redshift, then many CO emitters should show evidence for jets. Out of the 12 $z>3$ CO emitters\footnote{Five of the sources are likely to be gravitationally lensed, with the potential for distortion and differential magnification caused by the foreground lens. Therefore, although we include these sources in our discussion, we note that their observations cannot be so easily interpreted.}, six have evidence for jets aligned with either the CO or dust, five have radio luminosities above $10^{27}$~W~Hz$^{-1}$ and are clearly AGNs, and a further four have radio luminosities above $10^{25}$~W~Hz$^{-1}$ indicating either extreme starbursts or possible AGN. 
Figure~\ref{image:histo} shows a histogram of relative position angles for the subset of CO emitters with resolved radio and submm emission. Notwithstanding the small numbers, the clear excess centred around PA$\sim0^{\circ}$ is consistent with an alignment effect in $z>3$ CO emitters. We note that in the absence of radio jets, an alignment effect is expected for star forming galaxies since the radio emission will be tracing the same physical regions as the gas and dust. However, all six sources considered in this histogram have AGN and at least three are radio galaxies with detectable relativistic jets. In addition, there are many reasons why a jet may not be detected (or detectable) at radio wavelengths including (i) the active jet lifetime may be shorter than the period of continued star formation along the jet direction, (ii) radio observations made at rest frame centimetre wavelengths rather than (deci)metre wavelengths where lobes and hotspots are brighter, (iii) smaller magnetic fields or relativistic bulk flow with no relative motion between the electrons and the magnetic field, and (iv) rapid quenching of the radio emission due to IC losses against the CMB. Finally, since the CO emitters are heterogeneously selected (Table~1), they may not necessarily represent typical star-forming regions in the early universe. 
\section{Discussion}
\subsection{The radio-optical alignment effect}
\label{disc1}
There is tantalising observational evidence which supports jet-induced star formation as the natural explanation for the radio-optical alignment effect discussed in \S\ref{intro}. Well known examples exist in the local universe \citep{wvb85,rej02}, but the evidence is most conspicuous in high redshift radio galaxies like 4C~41.17 at $z=3.8$ where unpolarised, rest-frame UV continuum is aligned along the radio axis \citep{cha96,dey97}. If high redshift star formation is first triggered on large scales due to the above process, then the first molecular gas and dust to form will be located in those regions where the first stars expelled their enriched material into the inter-stellar medium (ISM). Therefore, we predict that if CO is detected in 4C~41.17, it will be aligned along the radio axis.
\subsection{In the beginning there were jets...}
\label{inthebegining}
Although the initial formation mechanisms of supermassive black holes remains largely unknown, the notion of seed black holes which form primordially and grow into a distribution of black hole masses has been around for three decades \citep{car74,sil98}. The mass distribution would necessarily be governed, at least partially, by the density of the surrounding gas; the most massive black holes would then form in regions of the highest gas density and it will be these sites where we observe high redshift radio galaxies and radio-loud quasars. The highly relativistic, supersonic jets which power into the surrounding medium and slam into the existing overdensities can trigger star formation along cocoons surrounding the jets \citep{bic00,fra04} or could modify the stellar initial mass function due to the effect of enhanced cosmic ray ionisation in the molecular cores (RDE; in preparation). Therefore, it will also be along these same preferential directions that the first heavy elements--- including carbon and oxygen ---will be produced as the stars end their lives and enrich their surroundings. This model provides the means of orchestrating star formation over tens of kpc on light crossing timescales. It is this process which has already been invoked to explain the radio-optical alignment effect at high redshift \citep{ree89}. Here we extend the idea to the formation of the first metals and molecules and predict, therefore, that a strong alignment exists generally between high-redshift molecular gas, dust and jets from supermassive black holes. \newline
  But are the timescales consistent? Existing models for jet-induced star formation predict cloud collapse to occur on timescales of order $10^6$ years. There will be a delay before CO formation takes place which, on the simplest level, will be the time it takes for the enrichment of the ISM. For 4C~41.17 (see \S\ref{disc1}), the age of the stellar population aligned with the radio axis is a few $\times10^6$ years. The lifetime of radio jets is typically a few $\times10^7$ years so multiple generations of stars can form along the axes of relativistic jets and consequently the CO to \htwo~abundance will increase to values approaching that seen in nearby galaxies. Once there are enough metals, molecules and dust available, then cooling and conventional star formation would take over, and be the dominant process in the local universe. \newline
Our hypothesis that AGN activity and jet formation precede the assembly of their host galaxy has wide ramifications. The continuum radiation from the jets could provide the flux needed to observe H\textsc{i} in absorption beyond the epoch of reionization \citep{car02c}. A jet-induced star formation scenario could also have profound implications for galaxy formation models. Should we be considering a paradigm in which the presence of such jets plays a role in deciding the morphology of the host galaxy?

\acknowledgments
We thank C. Carilli and E. Hu for the images of BR~1202$-$0725, D. Alexander for valuable discussions and the anonymous refereee for useful comments which improved this Letter. EMS and RWH acknowledge grant support from the Australian Research Council.

%% thebibliography produces citations in the text using \bibitem-\citep
%% to the KEY in the \citep commands (see the first section above).
%% place of the \citep commands.



%\clearpage

%\bibliography{mnemonic-simple,bibliography}

\begin{thebibliography}{}

\bibitem[\protect\citeauthoryear{{Barvainis}, {Alloin} \& {Bremer}}{{Barvainis}
  et~al.}{2002}]{bar02a}
{Barvainis} R.,  {Alloin} D., \&  {Bremer} M.  2002, A\&A, 385, 399

\bibitem[\protect\citeauthoryear{{Bertoldi}, {Carilli}, {Cox}, {Fan},
  {Strauss}, {Beelen}, {Omont} \& {Zylka}}{{Bertoldi} et~al.}{2003}]{ber03a}
{Bertoldi} F.,  {Carilli} C.~L.,  {Cox} P.,  {Fan} X.,  {Strauss} M.~A.,
  {Beelen} A.,  {Omont} A., \&    {Zylka} R.  2003, A\&A, 406, L55

\bibitem[\protect\citeauthoryear{{Bicknell}, {Sutherland}, {van Breugel},
  {Dopita}, {Dey} \& {Miley}}{{Bicknell} et~al.}{2000}]{bic00}
{Bicknell} G.~V.,  {Sutherland} R.~S.,  {van Breugel} W.~J.~M.,  {Dopita}
  M.~A.,  {Dey} A.,   \&  {Miley} G.~K.  2000, ApJ, 540, 678

\bibitem[\protect\citeauthoryear{{Bromm} \& {Larson}}{{Bromm} \&
  {Larson}}{2004}]{bro04}
{Bromm} V., \&  {Larson} R.~B.  2004, ARA\&A, 42, 79

\bibitem[\protect\citeauthoryear{{Carilli}, {Bertoldi}, {Omont}, {Cox},
  {McMahon} \& {Isaak}}{{Carilli} et~al.}{2001}]{car01}
{Carilli} C.~L.,  {Bertoldi} F.,  {Omont} A.,  {Cox} P.,  {McMahon} R.~G.,
\&   {Isaak} K.~G.  2001, AJ, 122, 1679

\bibitem[\protect\citeauthoryear{{Carilli}, {Kohno}, {Kawabe}, {Ohta},
  {Henkel}, {Menten}, {Yun}, {Petric} \& {Tutui}}{{Carilli}
  et~al.}{2002a}]{car02b}
{Carilli} C.~L., et al. 2002a, AJ, 123, 1838 (C02)

\bibitem[\protect\citeauthoryear{{Carilli}, {Gnedin} \& {Owen}}{{Carilli}
  et~al.}{2002b}]{car02c}
{Carilli} C.~L.,  {Gnedin} N.~Y.,  \&   {Owen} F.  2002b, ApJ, 577, 22

\bibitem[\protect\citeauthoryear{{Carilli}, {Lewis}, {Djorgovski}, {Mahabal},
  {Cox}, {Bertoldi} \& {Omont}}{{Carilli} et~al.}{2003}]{car03}
{Carilli} C.~L.,  {Lewis} G.~F.,  {Djorgovski} S.~G.,  {Mahabal} A.,  {Cox} P.,
   {Bertoldi} F., \&    {Omont} A.  2003, Science, 300, 773

\bibitem[\protect\citeauthoryear{{Carilli}, {Walter}, {Bertoldi}, {Menten},
  {Fan}, {Lewis}, {Strauss}, {Cox}, {Beelen}, {Omont} \& {Mohan}}{{Carilli}
  et~al.}{2004}]{car04}
{Carilli} C.~L., et al.  2004, {AJ in press }

\bibitem[\protect\citeauthoryear{{Carr} \& {Hawking}}{{Carr} \&
  {Hawking}}{1974}]{car74}
{Carr} B.~J., \&  {Hawking} S.~W.  1974, MNRAS, 168, 399

\bibitem[\protect\citeauthoryear{{Chambers}, {Miley} \& {van
  Breugel}}{{Chambers} et~al.}{1987}]{cha87}
{Chambers} K.~C.,  {Miley} G.~K.,   \&  {van Breugel} W.  1987, Nature, 329, 604

\bibitem[\protect\citeauthoryear{{Chambers}, {Miley}, {van Breugel}, {Bremer},
  {Huang} \& {Trentham}}{{Chambers} et~al.}{1996}]{cha96}
{Chambers} K.~C.,  {Miley} G.~K.,  {van Breugel} W.~J.~M.,  {Bremer} M.~A.~R.,
  {Huang} J.-S., \&    {Trentham} N.~A.  1996, ApJS, 106, 247

\bibitem[\protect\citeauthoryear{{Condon}}{{Condon}}{1983}]{con83}
{Condon} J.~J.  1983, ApJS, 53, 459

\bibitem[\protect\citeauthoryear{{Condon}}{{Condon}}{1992}]{con92}
{Condon} J.~J.  1992, ARA\&A, 30, 575

\bibitem[\protect\citeauthoryear{{Cowie}, {Barger} \& {Kneib}}{{Cowie}
  et~al.}{2002}]{cow02}
{Cowie} L.~L.,  {Barger} A.~J., \&    {Kneib} J.-P.  2002, AJ, 123, 2197

\bibitem[\protect\citeauthoryear{{Cox}, {Omont}, {Djorgovski}, {Bertoldi},
  {Pety}, {Carilli}, {Isaak}, {Beelen}, {McMahon} \& {Castro}}{{Cox}
  et~al.}{2002}]{cox02}
{Cox} P., et al. 2002, A\&A, 387, 406

\bibitem[\protect\citeauthoryear{{De Breuck}, {van Breugel}, {R{\" o}ttgering},
  {Stern}, {Miley}, {de Vries}, {Stanford}, {Kurk} \& {Overzier}}{{De Breuck}
  et~al.}{2001}]{cdb01}
{De Breuck} C. et al. 2001, AJ, 121, 1241

\bibitem[\protect\citeauthoryear{{De Breuck}, {Neri} \& {Omont}}{{De Breuck}
  et~al.}{2003a}]{cdb03a}
{De Breuck} C.,  {Neri} R.,   \&  {Omont} A.  2003a, New Astron. Rev, 47, 285


\bibitem[\protect\citeauthoryear{{De Breuck}, {Neri}, {Morganti}, {Omont},
  {Rocca-Volmerange}, {Stern}, {Reuland}, {van Breugel}, {R{\" o}ttgering},
  {Stanford}, {Spinrad}, {Vigotti} \& {Wright}}{{De Breuck}
  et~al.}{2003b}]{cdb03b}
{De Breuck} C., et al. 2003b, A\&A, 401, 911


\bibitem[\protect\citeauthoryear{{Dey}, {van Breugel}, {Vacca} \&
  {Antonucci}}{{Dey} et~al.}{1997}]{dey97}
{Dey} A.,  {van Breugel} W.,  {Vacca} W.~D., \&    {Antonucci} R.  1997, ApJ,
  490, 698

\bibitem[\protect\citeauthoryear{{Downes}, {Neri}, {Wiklind}, {Wilner} \&
  {Shaver}}{{Downes} et~al.}{1999}]{dow99}
{Downes} D.,  {Neri} R.,  {Wiklind} T.,  {Wilner} D.~J.,  \&   {Shaver} P.~A.
  1999, ApJ, 513, L1

\bibitem[\protect\citeauthoryear{{Fanaroff} \& {Riley}}{{Fanaroff} \&
  {Riley}}{1974}]{fan74}
{Fanaroff} B.~L., \&  {Riley} J.~M.  1974, MNRAS, 167, 31

\bibitem[\protect\citeauthoryear{{Fontana}, {D'Odorico}, {Giallongo},
  {Cristiani}, {Monnet} \& {Petitjean}}{{Fontana} et~al.}{1998}]{fon98}
{Fontana} A.,  {D'Odorico} S.,  {Giallongo} E.,  {Cristiani} S.,  {Monnet} G.,
  \&   {Petitjean} P.  1998, AJ, 115, 1225

\bibitem[\protect\citeauthoryear{{Fragile}, {Murray}, {Anninos} \& {van
  Breugel}}{{Fragile} et~al.}{2004}]{fra04}
{Fragile} P.~C.,  {Murray} S.~D.,  {Anninos} P., \&    {van Breugel} W.  2004,
  ApJ, 604, 74

\bibitem[\protect\citeauthoryear{{Greve}, {Ivison} \& {Papadopoulos}}{{Greve}
  et~al.}{2004}]{gre04}
{Greve} T.~R.,  {Ivison} R.~J., \&    {Papadopoulos} P.~P.  2004, A\&A, 419, 99


\bibitem[\protect\citeauthoryear{{Guilloteau}, {Omont}, {McMahon}, {Cox} \&
  {Petitjean}}{{Guilloteau} et~al.}{1997}]{gui97}
{Guilloteau} S.,  {Omont} A.,  {McMahon} R.~G.,  {Cox} P., \&    {Petitjean} P.
  1997, A\&A, 328, L1

\bibitem[\protect\citeauthoryear{{Guilloteau}, {Omont}, {Cox}, {McMahon} \&
  {Petitjean}}{{Guilloteau} et~al.}{1999}]{gui99}
{Guilloteau} S.,  {Omont} A.,  {Cox} P.,  {McMahon} R.~G., \&    {Petitjean} P.
  1999, A\&A, 349, 363

\bibitem[\protect\citeauthoryear{{Hu}, {McMahon} \& {Egami}}{{Hu}
  et~al.}{1996}]{hu96}
{Hu} E.~M.,  {McMahon} R.~G., \&    {Egami} E.  1996, ApJ, 459, L53 (H96)

\bibitem[\protect\citeauthoryear{{Ibata}, {Lewis}, {Irwin}, {Leh{\' a}r} \&
  {Totten}}{{Ibata} et~al.}{1999}]{iba99}
{Ibata} R.~A.,  {Lewis} G.~F.,  {Irwin} M.~J.,  {Leh{\' a}r} J.,  \&   {Totten}
  E.~J.  1999, AJ, 118, 1922

\bibitem[\protect\citeauthoryear{{Ivison}, {Greve}, {Smail}, {Dunlop}, {Roche},
  {Scott}, {Page}, {Stevens}, {Almaini}, {Blain}, {Willott}, {Fox}, {Gilbank},
  {Serjeant} \& {Hughes}}{{Ivison} et~al.}{2002}]{ivi02}
{Ivison} R.~J., et al. 2002, MNRAS, 337, 1

\bibitem[\protect\citeauthoryear{{Kauffmann}, {White} \&
  {Guiderdoni}}{{Kauffmann} et~al.}{1993}]{kau93}
{Kauffmann} G.,  {White} S.~D.~M.,  \&   {Guiderdoni} B.  1993, MNRAS, 264, 201

\bibitem[\protect\citeauthoryear{{Lehar}, {Burke}, {Conner}, {Falco},
  {Fletcher}, {Irwin}, {McMahon}, {Muslow} \& {Schechter}}{{Lehar}
  et~al.}{1997}]{leh97}
{Lehar} J., et al. 1997, AJ, 114, 48

\bibitem[\protect\citeauthoryear{{McCarthy}, {van Breugel}, {Spinrad} \&
  {Djorgovski}}{{McCarthy} et~al.}{1987}]{mcc87}
{McCarthy} P.~J.,  {van Breugel} W.,  {Spinrad} H., \& {Djorgovski} S.  1987,
  ApJ, 321, L29

\bibitem[\protect\citeauthoryear{{McMahon}, {Omont}, {Bergeron}, {Kreysa} \&
  {Haslam}}{{McMahon} et~al.}{1994}]{mcm94}
{McMahon} R.~G.,  {Omont} A.,  {Bergeron} J.,  {Kreysa} E., \& {Haslam}
  C.~G.~T.,  1994, MNRAS, 267, L9

\bibitem[\protect\citeauthoryear{{Neri}, {Genzel}, {Ivison}, {Bertoldi},
  {Blain}, {Chapman}, {Cox}, {Greve}, {Omont} \& {Frayer}}{{Neri}
  et~al.}{2003}]{ner03}
{Neri} R., et al. 2003, ApJ, 597, L113

\bibitem[\protect\citeauthoryear{{Ohta}, {Yamada}, {Nakanishi}, {Kohno},
  {Akiyama} \& {Kawabe}}{{Ohta} et~al.}{1996}]{oht96}
{Ohta} K.,  {Yamada} T.,  {Nakanishi} K.,  {Kohno} K.,  {Akiyama} M.,
  \& {Kawabe} R.  1996, Nature, 382, 426

\bibitem[\protect\citeauthoryear{{Omont}, {Petitjean}, {Guilloteau}, {McMahon},
  {Solomon} \& {Pecontal}}{{Omont} et~al.}{1996}]{omo96a}
{Omont} A.,  {Petitjean} P.,  {Guilloteau} S.,  {McMahon} R.~G.,  {Solomon}
  P.~M.,  \&   {Pecontal} E.  1996, Nature, 382, 428


\bibitem[\protect\citeauthoryear{{Papadopoulos}, {R{\" o}ttgering}, {van der
  Werf}, {Guilloteau}, {Omont}, {van Breugel} \& {Tilanus}}{{Papadopoulos}
  et~al.}{2000}]{pap00}
{Papadopoulos} P.~P.,  {R{\" o}ttgering} H.~J.~A.,  {van der Werf} P.~P.,
  {Guilloteau} S.,  {Omont} A.,  {van Breugel} W.~J.~M., \&    {Tilanus} R.~P.~J.
   2000, ApJ, 528, 626

\bibitem[\protect\citeauthoryear{{Papadopoulos}, {Ivison}, {Carilli} \&
  {Lewis}}{{Papadopoulos} et~al.}{2001}]{pap01}
{Papadopoulos} P.,  {Ivison} R.,  {Carilli} C.,  \&   {Lewis} G.  2001, Nature,
  409, 58

\bibitem[\protect\citeauthoryear{{Rees}}{{Rees}}{1989}]{ree89}
{Rees} M.~J.  1989, MNRAS, 239, 1

\bibitem[\protect\citeauthoryear{{Rejkuba}, {Minniti}, {Courbin} \&
  {Silva}}{{Rejkuba} et~al.}{2002}]{rej02}
{Rejkuba} M.,  {Minniti} D.,  {Courbin} F., \&    {Silva} D.~R.  2002, ApJ, 564,
  688

\bibitem[\protect\citeauthoryear{{Silk} \& {Rees}}{{Silk} \&
  {Rees}}{1998}]{sil98}
{Silk} J.,  {Rees} M.~J.  1998, A\&A, 331, L1

\bibitem[\protect\citeauthoryear{{Stevens}, {Ivison}, {Dunlop}, {Smail},
  {Percival}, {Hughes}, {R{\" o}ttgering}, {van Breugel} \&
  {Reuland}}{{Stevens} et~al.}{2003}]{ste03}
{Stevens} J.~A., et al. 2003, Nature, 425, 264

\bibitem[\protect\citeauthoryear{{Storrie-Lombardi}, {McMahon}, {Irwin} \&
  {Hazard}}{{Storrie-Lombardi} et~al.}{1996}]{sto96}
{Storrie-Lombardi} L.~J.,  {McMahon} R.~G.,  {Irwin} M.~J.,  \&   {Hazard} C.
  1996, ApJ, 468, 121

\bibitem[\protect\citeauthoryear{{van Breugel}, {Filippenko}, {Heckman} \&
  {Miley}}{{van Breugel} et~al.}{1985}]{wvb85}
{van Breugel} W.,  {Filippenko} A.~V.,  {Heckman} T., \&    {Miley} G.  1985,
  ApJ, 293, 83

\bibitem[\protect\citeauthoryear{{van Breugel}, {Stanford}, {Spinrad}, {Stern}
  \& {Graham}}{{van Breugel} et~al.}{1998}]{wvb98}
{van Breugel} W.~J.~M.,  {Stanford} S.~A.,  {Spinrad} H.,  {Stern} D., \& 
  {Graham} J.~R.  1998, ApJ, 502, 614

\bibitem[\protect\citeauthoryear{{Walter}, {Bertoldi}, {Carilli}, {Cox}, {Lo},
  {Neri}, {Fan}, {Omont}, {Strauss} \& {Menten}}{{Walter} et~al.}{2003}]{wal03}
{Walter} F., et al. 2003, Nature, 424, 406

\bibitem[\protect\citeauthoryear{{White} \& {Rees}}{{White} \&
  {Rees}}{1978}]{whi78}
{White} S.~D.~M., \&   {Rees} M.~J.  1978, MNRAS, 183, 341


\bibitem[\protect\citeauthoryear{{Willott}, {Rawlings}, {Blundell}, {Lacy} \&
  {Eales}}{{Willott} et~al.}{2001}]{wil01}
{Willott} C.~J.,  {Rawlings} S.,  {Blundell} K.~M.,  {Lacy} M.,  \&   {Eales}
  S.~A.  2001, MNRAS, 322, 536

\bibitem[\protect\citeauthoryear{{Willott}, {McLure} \& {Jarvis}}{{Willott}
  et~al.}{2003}]{wil03}
{Willott} C.~J.,  {McLure} R.~J., \&    {Jarvis} M.~J.  2003, ApJ, 587, L15

\bibitem[\protect\citeauthoryear{{Yun}, {Carilli}, {Kawabe}, {Tutui}, {Kohno}
  \& {Ohta}}{{Yun} et~al.}{2000}]{yun00}
{Yun} M.~S.,  {Carilli} C.~L.,  {Kawabe} R.,  {Tutui} Y.,  {Kohno} K.,
\&   {Ohta} K.  2000, ApJ, 528, 171 

\end{thebibliography}
\begin{deluxetable}{rllcllll}
\tabletypesize{\scriptsize}

\tablecaption{Radio, CO and submm properties of all published $z>3$ CO emitters.\label{tbl-1}}
\tablewidth{0pt}
\tablehead{
\colhead{Source} & \colhead{$z_{opt}$} & \colhead{$z_{\textsc{co}}$} & \colhead{Radio PA}\tablenotemark{a}\tablenotemark{b} & \colhead{Radio Power}\tablenotemark{a} & \colhead{CO PA}\tablenotemark{b} & \colhead{Dust PA}\tablenotemark{b} & \colhead{References} \\
\colhead{} & \colhead{} & \colhead{} & \colhead{$^{\circ}$} & \colhead{W~Hz$^{-1}$} & \colhead{$^{\circ}$} & \colhead{$^{\circ}$} & \colhead{} }
\startdata

SDSS~J1148+5251\tablenotemark{c} &  6.41 & 6.42 &\nodata&$1\times10^{25}$\tablenotemark{f}  &\nodata &\nodata &  1,2,3,4\\% \citealp{wil03,wal03,ber03a,car04} 1,2,3,4\\
BR~1202$-$0725\tablenotemark{c} & 4.694 & 4.695 & $-$35 & $3\times10^{25}$ &  $-$35  & $-$40  & 5,6,7\\% \citealp{sto96,omo96a,car02b}\\%5,6,7
BRI~0952$-$0115\tablenotemark{c}\tablenotemark{d} & 4.426 & 4.434 &\nodata &$4\times10^{25}$\tablenotemark{g} &\nodata&\nodata &5,8,9 \\% \citealp{sto96,gui97}\\%5,8,9 
BRI~1335$-$0417\tablenotemark{c} & 4.396 & 4.407 &\nodata & $3\times10^{25}$ &  +13 & +18 & 5,8,9,7\\% \citealp{sto96,gui97,gui99,car02b}\\%5,8,9,7
PSS~J2322+1944\tablenotemark{d} & 4.111 & 4.119 & 0 &$4\times10^{24}$\tablenotemark{g} &  +5 &\nodata&10,11,12\\% \citealp{car01,car03,cox02}\\%10,11,12
APM~08279+5255\tablenotemark{d} & 3.87 & 3.91 & +23 &$1\times10^{26}$\tablenotemark{g} &  +20 &\nodata&13,14,15\\% \citealp{iba99,pap01,dow99}\\%13,14,15
4C~60.07\tablenotemark{e} & 3.788 & 3.791 & $-$73  &$4\times10^{28}$ &  $-$75:\tablenotemark{h}  &  $-$70 & 16,17,18,19\\%\citealp{cha96,pap00,ste03,gre04}\\%16,17,18,19
6C~1909+722\tablenotemark{e} & 3.536 & 3.532 & +15 & $5\times10^{28}$ &\nodata& +18 & 20,17,18\\%\citealp{cdb01,pap00,ste03}\\%20,17,18
TN~J0121+1320\tablenotemark{e} & 3.516 & 3.520 &\nodata&$1\times10^{28}$  & \nodata&\nodata& 20,21\\%\citealp{cdb01,cdb03a}\\%20,21
SMM~09431+4700\tablenotemark{d} & 3.35 & 3.35 &\nodata & $4\times10^{24}$\tablenotemark{f}\tablenotemark{g} & \nodata&\nodata&22,23\\% \citealp{cow02,ner03}\\%22,23
B0751+2716\tablenotemark{d}\tablenotemark{e}  & 3.200 &3.200 &\nodata &$1\times10^{28}$\tablenotemark{g} &\nodata&\nodata & 24,25\\%\citealp{bar02a,leh97}\\%24,25
B3~J2330+3927\tablenotemark{e}  & 3.087 & 3.094 & +33 & $9\times10^{27}$& \nodata&  +50 &   26,18%\citealp{cdb03b,ste03} \\%26,18
\enddata
%% Text for table notes should follow after the \enddata but before
%% the \end{deluxetable}. Make sure there is at least one \tablenotemark
%% in the table for each \tablenotetext.

\tablecomments{$^{\textrm{a}}$Radio luminosity measured at 1.4GHz in the rest frame of the source. $^{\textrm{b}}$PAs are measured from published images given in the reference list. Errors in PA are of order $5^{\circ}$. $^{\textrm{c}}$Optically selected quasar with submm emission. $^{\textrm{d}}$Gravitationally lensed with submm emission. $^{\textrm{e}}$Radio galaxy with submm emission. $^{\textrm{f}}$Assuming a radio spectral index of $-$0.7. $^{\textrm{g}}$Not corrected for possible gravitational magnification. $^{\textrm{h}}$Based on the broad CO(1-0) component of \citet{gre04}.}

%\tablenotetext{a}{Radio luminosity measured at 1.4GHz in the rest frame of the source. }
%\tablenotetext{b}{PAs are measured from published images given in the reference list. Errors in PA are of order 10\%}
%\tablenotetext{c}{Optically selected quasar with submm emission}
%\tablenotetext{d}{Gravitationally lensed with submm emission}
%\tablenotetext{e}{Radio galaxy with submm emission}
%\tablenotetext{f}{Assuming a radio spectral index of $-$0.7}
%\tablenotetext{g}{Not corrected for possible gravitational magnification}
\tablerefs{
(1) \citealp{wil03} (2) \citealp{wal03} (3) \citealp{ber03a} (4) \citealp{car04} (5) \citealp{sto96} (6) \citealp{omo96a}
(7) \citealp{car02b} (8) \citealp{gui97} (9) \citealp{gui99} (10) \citealp{car01} (11) \citealp{car03} (12) \citealp{cox02}
(13) \citealp{iba99} (14) \citealp{pap01} (15) \citealp{dow99} (16) \citealp{cha96} (17) \citealp{pap00} (18) \citealp{ste03}
(19) \citealp{gre04} (20) \citealp{cdb01} (21) \citealp{cdb03a} (22) \citealp{cow02} (23) \citealp{ner03} (24) \citealp{bar02a}
(25) \citealp{leh97} (26) \citealp{cdb03b}
}

\end{deluxetable}
\clearpage

%\clearpage

%\clearpage

\end{document}